# E-Health Initiative and Customer's Expectation: Case Brunei

Mohammad Nabil Almunawar [1], Zaw Wint[2], Patrick Kim Cheng Low[3], Muhammad Anshari [4]

*Abstract*— This paper is to determine the dimension of e-health services in Brunei Darussalam (Brunei) from customers' perspective. It is to identify, understand, analyze and evaluate public's expectation on e-health in Brunei. A questionnaire was designed to gather quantitative and qualitative data to survey patients, patient's family, and health practitioners at hospitals, clinics, or home care centers in Brunei starting from February to March, 2011. A 25-item Likert-type survey instrument was specifically developed for this study and administered to a sample of 366 patients. The data were analyzed to provide initial ideas and recommendation to policy makers on how to move forward with the e-health initiative as a mean to improve healthcare services. The survey revealed that there exists a high demand and expectation from people in Brunei to have better healthcare services accessible through an e-health system in order to improve health literacy as well as quality and efficiency of healthcare. Regardless of the limitations of the survey, the general public has responded with a great support for the capabilities of an e-health system listed from the questionnaires. The results of the survey provide a solid foundation for our on going research project to proceed further to develop a model of e-health and subsequently develop a system prototype that incorporate expectations from the people.

## I. INTRODUCTION

A successful organization exists when it provides good quality service and keeps offering more services [1]. The fact that customer service expectations in healthcare organization are high poses a serious challenge for healthcare providers as they have to make exceptional impression on every customer. Healthcare organizations strategies should transform customer strategies and systems to customer engagement. Proactive strategies will improve customer services [2].

The use of Information & Communication Technology (ICT) in healthcare organizations is not new. Deploying ICT in healthcare environment has helped healthcare professionals to improve the efficiency and effectiveness of services. Healthcare information systems that can record and locate important information quickly have become a standard practice in many healthcare organizations. Recently, issues in utilizing ICT in providing better healthcare service have become challenging research in response to the advancement of ICT. The use of ICT will provide healthcare organizations with a better understanding on features and services that can improve quality of services. ICT initiatives in a healthcare organization, known as e-health, provide many advantages such as cost efficiency and patients' empowerment. For instance, online consultation can offer more diverse audiences (patients) that unable to come physically for consultation. Additionally, patients' empowerment allows patients/patient's family to gain access to their medical records online which is good to create sense of awareness to the status their health. In addition, e-health extends the scope of conventional healthcare boundaries. Regardless the location, physicians can provide consultation over the Internet without even meeting their patients physically. Additionally, e-health offers to connect doctors and other healthcare professionals with their patients online [3].

In order to explore the expectation of patients for the benefits of e-health above, we have conducted a survey on e-health systems across the country in Brunei. The survey is designed to find out the demographical traits of potential users, their opinion of e-health services as well as disseminating the potential abilities of e-health to the public.

The objective of the survey is to assess the respondents' awareness and demand for e-health services in Brunei. The survey is also aim to expose online features needed by the system whether the implementation of e-health system will be helpful to the people, improve customer satisfaction and health literacy among the people which may lead to reduce economic burden in the long run for the society. Moreover, the result may contribute to portray the current state of demographical patterns and general public opinion on e-health initiatives.

The survey revealed important facts that Brunei is comprised of well-educated and highly skilled population especially in terms of applying ICT skills. The survey also depicted the opinion of people in Brunei that the majority of respondents favour to use services provided through e-health systems compared with services that offered through conventional healthcare systems. To draw the result of the survey, the structure of the paper is designed as follow. In sections 2, we discuss background of the study and recent status of customer service in healthcare organizations. The technical details and analysis of survey are discussed in Section 3 and 4 respectively, and finally we conclude our findings in form of recommendations for future direction in section 5.

## II. BACKGROUND OF STUDY

To manage customers, an organization needs to know how to deal with their expectations. This involves the ability to detect their expectations and to identify reasons of their

1.Mohammad Nabil Almunawar, 2.Zaw Wint, 3. Patrick Kim Cheng Low, and Muhammad Anshari ; FBEPS-UBD, Jl.Tungku Link Gadong, BE 1410 Brunei Darussalam. Tel.: + 673 8825711; fax: +673 2465017 . E-mail address respectively: nabil.almunawar@ubd.edu.bn, zaw.wint@ubd.edu.bn, patrick.low@ubd.edu.bn, and anshari@yahoo.com

dissatisfactions. The survey focused on customers' perceptions, in other words, this is an attempt to listening to customers in order to understand their needs and expectations.

### A. Healthcare Customer Service

Providing excellent services to customers is becoming vital in today's competitive environment. To survive in today's competitive market, healthcare organizations must incorporate customer's viewpoint [1]. Several researchers explained that improving quality of service has a positive effect on customer's satisfaction [4].

Customers' satisfactions can be considered as an integrated response to evaluation. In other words, improving customers' satisfaction should be a central notion to derive business strategy [5]. Also, offering premium quality of services has become the most powerful competitive weapon [6]. Since the past decades, involving customers in healthcare processes has been increasingly emphasized, with an appreciation of positive impact on outcomes that follows. As a business entity, a healthcare organization needs to practice the same standards of customer service as practiced by other industries or business organizations.

Nowadays, customers in healthcare are actively seeking health information and using it to make decisions about their health. The high expectations in service on healthcare organizations pose a serious challenge for healthcare organizations to fulfill as they have to make exceptional impression on every customer. As a matter of fact, poor service may lead customers to switch healthcare providers because poor service indicates inefficiency, higher cost and lower quality of care. Entel et.al [7] recommends that healthcare organizations should become "empathy engines," that is transforming their organizations to allow frontline employees to focus on patient problems and innovate. This applies to hospitals, clinics, payers, vendors and pharmacy chains as well.

Customer service is not an "extra" thing. It is an essential requirement for providing high quality healthcare and for staying in highly competitive business [8]. Patients are making clear choices about where they receive care based on service experiences; therefore it is crucial for organizations to create an institutional ability to sense and respond empathetically [7]. Nowadays patients have more choices to seek care and interact with their healthcare providers. A great customer service can lead to major improvements in the healthcare system. In addition, a deep understanding of the interaction among factors involved such as quality of healthcare services, their outcomes, and patient satisfaction, has become an invaluable input for designing and managing healthcare systems [9]. For this purpose, it is necessary to investigate dimensions service quality, which customers perceive as important, in the healthcare context [6].

### B. E-Health Systems

Nowadays, people at all ages, including children and elders, frequently access information on the Internet. ICT can be used to provide effective self service and empower customers' care that allows organization to reduce costs by handling an increasing number of customer transactions efficiently. Meeting customers' expectations will pose a significant challenge on how an organization attaches with their customers effectively. Thus, the ICT strategy should be aligned with the business strategy in which customers' satisfaction, in ways that acquire new customer, retain them to use the facilities, and extend them to use the other services, and must be addressed. Additionally, customers' demands have changed because customers felt empowered by the large amount of information available to them [10]. Empowerment on an individual level is a social process of recognizing, promoting, and enhancing peoples' abilities to meet their own needs, solve their own problems and mobilize the necessary resources in order to control of their lives [11]. There is good evidence that coaching patients using empowerment strategies leads to broadened, less negative definitions of illness as well as improved self management by patients. However, without commitment responses from healthcare organization, the benefits of empowerment are unlikely to emerge [12].

Following the fast pace of advancement in the Web technology, it is inevitable that healthcare service utilizes the Web technology to develop a more efficient and effective healthcare organization. The Web technology allows medical staffs to monitor their patients remotely and to share information conveniently on health education or promotion programs with their patients. Utilizing the Web for healthcare, which often refers to e-health, has several advantages such as availability, fast response and convenience.

With the advancement of ICT, Web 2.0 technology has brought a possibility to extend the service of e-health by enabling patients, patient's families, and community at large to participate more actively in the process of health promotion and education through social networking process.

In light of the Web technology applied on healthcare, Eysenbach defined e-health as an emerging field in the intersection of medical informatics, public health and business, referring to health services and information delivered or enhanced through the Internet and related technologies [13]. In a broader sense, the term characterizes not only a technical development, but also a state of mind, a way of thinking, an attitude, and a commitment for network, global thinking, to improve health care locally, regionally, and worldwide by using ICT.

E-health is gaining its popularity. There are 193 countries that are a member of World Health Organization (WHO) in 2009; 114 of them participated in the global survey on e-health [14]. Most developed countries have fully utilized e-health in their systems as they have the resources, expertise, and capital to implement them. While developing countries e-health systems have not been fully utilized yet.

Several examples of countries that have implemented e-health are Canada, Singapore and Australia to name a few. Canada has established e-health Ontario on March 2009 with three targeted strategies to improve; diabetes management, medication management and wait times. One of the examples of the service offered is ePrescribing under the medication management. It authorizes and transmits prescriptions from physicians and other prescribers to pharmacists and other dispensers [15]. It prevents medication error due to illegible prescribing and reduces fraud prescriptions. Participating prescribers and pharmacies at both sites will continue electronically prescribing until a provincial Medication Management System is in place [15].

*C. E-Health in Brunei*

According to a 2007 World Health Organization (WHO) global survey of health services, Brunei ranked 40th out of 190 countries and 2nd in the ASEAN nations and 4th within the Western Pacific region [16]. Furthermore, the Ministry of Health, Brunei in "National Healthcare 2000-2010: A strategic framework for action June 2000," stated that the Ministry's vision is to strive to become a highly reputable health service organization which is comparable to the best in the region and which will enable every citizen and resident of the nation to attain high quality of life for being socially, economically and mentally productive throughout the life span.

Meanwhile, its mission is to improve the health and well being of the people of Brunei through a highly quality and comprehensive health care system which is effective, efficient, responsive, affordable, equitable and accessible to all in the country. National Development Plan (NDP) was first implemented in 2007, where the budget of $149, 152, 000 is allocated for medical and health-care. The application of new technologies is included in this huge amount of money [17].

The statistic on Brunei Internet users from 2006 to 2009, has shown an increase in penetration of the Internet of 78.5% in Oxford Business Group's 2010 report on Brunei. Brunei's network infrastructure will be upgraded to ultra–broadband service connection through fibre optics by late August 2011. The high speed broadband is finally coming to Brunei with a maximum data transfer rate of 150mbps (megabits per second), a more stable, better network coverage, and consistent Internet experience [18]. This figure brings a promise to the initiative of e-health in the country. Though, e-health has not fully implemented yet in Brunei. Both private and public health institutions in Brunei are still using the manual system. Future e-health services may be promoted by a variety of providers including health practitioners, public health entities, individuals, groups from private organization like health clubs. The implication of the study for health practitioners, customers, healthcare providers and agents of care, ICT developers, researchers, and policy makers are proposed.

III. METHODOLOGY

We use the purposive sampling methods in which respondents were intentionally selected from patients, patient's family, or medical staffs from hospitals, clinics, and homecare centre across the country. Majority of the respondents are local Bruneians, who frequently visit health practitioners. The respondents range from twenty years old (or younger) to above 51 years old. Therefore, they represent a fair share of the general public. There were 366 respondents participating for the survey which conducted from February to March 2011. The time taken to complete the questionnaire was between 4-15 minutes. Some respondents required as long as 15 minutes while others can finish it off in 4 minutes. This was mainly due to their difficulties in answering and understanding the questions.

There were several problems when conducting the survey. Language barrier was one of the reasons. There were few respondents who were unable to understand English, thus they need Malay translation. There were also some difficulties for elderly people to fill the questionnaire since they do not understand the meaning of e-health. Thus, a prior explanation needed to be done. This problem was not seen for younger adults, in which they understand the concept of e-health. This could be explained by the fact that the younger generations nowadays are used to with the Internet and electronic media.

The 25 questions in the questionnaires were tailored to the population for this study was divided into three sections. The first section included 6 questions about demographics' traits of respondents where we learnt on our respondent gender composition, age, employment type, educational level, computer literacy, and the time they spend on the Internet. Nineteen questions in the second part asked about features of e-health that were specifically related to their healthcare service. The third section was an open comment that was intended to collect respondents' opinions. The questions as well as comments were analyzed to explore the relationships among variables for future recommendations, especially for developing an e-health framework..

IV. RESULTS

Because no research has ever been published on e-health and health information accessibility issues in Brunei, this paper is prepared to fill that gap. Many e-health topics are covered, including the use of existing e-health service, health information accessibility, and current issues like social support and social networking. Data gathered from the survey will be used to formulate recommendation for future direction of e-health systems.

In order to know a coefficient of reliability (consistency) of questionnaire, the authors use Cronbach's alpha to measure of internal consistency, that is, how to closely related set of items are as a group.

$$\alpha = \frac{N.\bar{c}}{\bar{v} + (N-1).\bar{c}} \quad (1)$$

Here N is equal to the number of items, c-bar is the average inter-item covariance among the items and v-bar equals the average variance as in (1). If Cronbach's alpha is greater than or equal to 0.6 then a variable is reliable.

**TABLE 1**
COEFFICIENT OF RELIABILITY

| | | N | % |
|---|---|---|---|
| Cases | Valid | 363 | 99.2 |
| | Excluded | 3 | .8 |
| | Total | 366 | 100.0 |

From the table 1 shows that Cronbach's Alpha is found 0.80 from the 363 number of items. It signifies that they have relatively high internal consistency since it is 0.80 > 0.60, hence the survey result is reliable.

From the survey, we found that expectations in healthcare services are high, which create a serious challenge for healthcare organizations in Brunei. In response to the expectation, it is significant to add features as front end to tie up the interface of e-health. Therefore, it is important to improve customer satisfaction and health literacy in healthcare service to accommodate components and features of social networking capabilities, empowering patients, and availability of online health educator (see figure 1).

Demographic Information

The first section our survey questionnaire was designed to gather data about employment type, gender, age composition, living arrangements, level of education, frequency of computer usage, frequency of Internet usage. When it comes to working pattern of the country, it is imperative to shed light on the respondents work at the government sector, which is 58%, while the rest, 42%, work at the private sector. This reflects the employment pattern in Brunei, where more people work with the government and the new generation is also inclined to work with the government for a job security reason and welfare benefits provided. The data reveal that there were more females than males in the survey conducted. According to the results 54% of female respondents completed the survey in comparison to the male who has completed the survey 46% only.

While in the age distribution section, the highest respondents were ranged between the ages of 21-30 years old with the percentage of 38%. There were 19% respondents in the range of 31 to 40, 18% of them were in the ranged of 41 to 50, where 13 % of them ageing between 20 years old or younger, and 12% were 51 years old and above. This shows that the people at these ages 21 up to 30 years old are the most approachable, cooperative and their command in English is better than the remaining age group so they could understand the questions. Respondents also were asked on their living arrangements. Some of them either live alone, with family or others such as with friends or lodging. 5% of the respondents live on their own, 90% of them live with their families and 5% live with others. It shows that almost all of the respondents live with the families as this is a typical Bruneian tradition, regardless of their marital status.

In term of educational level, each respondent has different level of education. The difference in education level might reflect. 10% of the respondents did not complete high school, 31% of them had completed only high school and 59% have completed more than high school level. This interprets that there are more educated respondents than those with lower education level. The result of the level of education might reflect the result of other factors such as their understanding towards the modern technology especially in ICT they are interrelated.

Additionally, the respondents were questioned on how frequent they use the Internet, either at home or at their workplace, regardless of what medium they use to get internet access. Based on the survey, almost of the Bruneians regardless of their age can easily have access to the Internet, 63% of the respondents use the Internet at least daily, 18% uses less than daily to weekly, 9% uses the Internet between weekly and monthly, and 10%, has never used Internet. The result shows that the majority of the respondents are exposed to the Internet and they tend to use the Internet everyday.

The respondents were also asked on how frequent they use computer in their everyday life, regardless of the purposes. The majority of the respondents (64% of them) use the computer everyday, 18% uses computer less than daily to weekly, 9% of them uses computer weekly to monthly, and 9% of them has never use the computer at all. This shows that the computer literacy in Brunei is high and this statement supports the results for usage of Internet and educational level as well.

Accessibility of Health Information

In this section, the questionnaires concentrate on accessibility of information to patient online. The first question was asking their agreement to make appointment online with the healthcare organization. Most of the respondents (83%) prefer to be able to make an appointment online. If the online appointment available, it reduces the waiting time for the patients to set the available day and time for their appointment so this will give patient convenience in service and efficiency in time.

When the respondents were asked their opinion on the ability of service to view their medical records online, the study shows that 33% (strongly agree) and 46% (agree) of the respondents prefer to view and have control to their medical records online. This is an interesting result which indicates that they can self-monitor their medical record. The knowledge of medical status from historical data available in the medical records may lead to improve healthcare awareness and self managed healthcare. In addition, the online medical record may help them to check and make sure that they have the right health details to avoid miscommunications. Further, the online medical records can be used as a guideline in making any decision relating to their own health. Although most of the respondents prefer online medical records, some respondents (17%) disagree with idea of their medical records accessible online. This is mainly due to the fear of information breach or manipulation. It is a challenge for future research in e-health implementation, the security and privacy issues need to be considered seriously.

The preference of online accessibility and communication is further enhanced by the tendency to prefer online referral. There are 78% of respondents agree to have an ability for requesting referral online as it facilitates patients to request their preferred or trustworthy referral online. An interesting finding also was encountered when respondents were asked about their ability to view their medical/healthcare bills. The survey reveals that 84% of the respondents like the service to view their medical bills and insurance coverage details online to keep them up-to-date on their payment information.

Again, the majority of the respondents prefer online service due to time efficiency. The survey reveals that 25% of the respondents strongly agreed to view and interact with the health promotion program online, 56% of them agreed, 16% disagreed, and 3% has strongly disagreed. For example, when patients are assigned to have a dietary program, they do not have to travel all the way to the hospital because they can have access the program using the online service provided and it saves time and money. Another benefit also for health professionals, they can monitor their patients' progress online, which is convenient for both parties. In addition, patients can easily share their health status after conducting their health program to their friend and relative which is good to gain moral support as well as to promote others to participate.

However, when they were asked the possibility to pay any of online service, half of the respondents would rather not to pay for any online service. The reason is most likely because most Bruneians are already comfortable with the free services provided by the government.

When we asked about their possibilities to communicate and consult with health educator online, most of the respondents (25% strongly agree and another 48% of them agree) prefer to communicate with their health promoter or health educator online. Online communication is faster and recorded, furthermore some people are more extrovert and more comfortable when communicating virtually.

During daily health life cycle, we asked them if they were able to record of their health related activities online. Most respondents were happy if they can record their health-related activities online (20% strongly and 56% agree) these activities may include a personal health diary where the respondents can access it, anytime and anywhere, facilitating their daily plans and programs for a healthy lifestyle. This service can be used to monitor their health status and may help in making health decisions later.

When respondents were asked the possibility to share anything related to health services provided from a hospital with others in social networks, the majority of them (45% agree and 18 % strongly agree) support such a sharing. They may think that sharing is a good way to improve health service - since they can discuss certain issues.

Furthermore, we asked their opinion on sharing health problem or experiences online with those who have similar problems, the survey reveals that most of the respondents (56% agree and 21% strongly agree) that they wanted to get in touch with other patients who have the same condition online. People with similar problems can easily share their experience and knowledge online which leads to supporting each other at least morally in facing their problems.

However, when they were asked what if third parties (pharmaceutical/insurance company/medical business, universities, etc) want to get access to their medical records either for research development or health product promotion, most respondents (about 80% of them) did not want pharmaceutical/insurance company/medical business entities to have access to their personal medical records online, unless there is prior consent from the particular individual.

Final section of the survey allows respondents to comment freely on their expectations of an e-health system and its accessibility in providing better knowledge on health status. Majority of those who filled in their comments support the list of abilities provided by an e-health system such as convenience in service, time saving, health promotion, accessibility, aware of medical result better, supporting each other, etc. For instant, a patient faced the daunting tasks of travelling, there is significant time spent in finding parking space, time spent in the waiting room and then finally waiting for the prescription medicines. So by using ICT, a patient can communicate with the medical healthcare professional from the comfort of his or her home. Through the health promotion program, social support systems, e-appointment, online consultation, and all features offered from the e-health systems expected that it will reduce the number of potential patients visit the hospital significantly, improve the quality of consultation, more productive time to other activities, and finally empowering people to responsible (self health management) to improve their health status. Furthermore, support groups such as social networks in healthcare in Brunei are not common. Respondents stated their endorsement on the possibility of support groups. A patient with who might have an illness that is socially shunned could find refuge in a social group with other patients who are condition as they are.

## V. DISCUSSION AND RECOMENDATIONS

In the survey we defined four key enablers and critical success factors that provide direction, resource to initiate, and realize their potential disposition from customer perspective. Understanding and managing of critical success factors are critical for a successful of an e-health initiative.

The research points to the issue of demographical pattern such as age composition, education levels, computer literacy, and Internet literacy, as enabler factors that make up typical e-health initiatives. It is important to see how the enablers have affected the preference of e-health services offered. The finding of this survey examines each of the four enablers and provides insight into their relative impact on e-health systems, and it is followed some recommendations based on the survey findings.

Enabler Factors

Level of education will affect the acceptability of systems and competency of people to utilize it. Higher educational level means higher possibility to success. In case of Brunei that the education is free and available for everyone, half of the respondents have completed high school, confirmed those who are literate enough understand easier about e-health systems when it is implemented. While comparing gender composition, It is encouraging to note that the female more than male in healthcare activities. This trend could because of females have more time to visit healthcare for treatment or they are more vulnerable in health matters than male. Age composition; majority of respondents are young (from 21 to 50 years group age). This indicates that younger people are likely to become potential users if it is implemented since they more expose to the technology. Internet accessibility; Majority of the respondents are exposed to the Internet and they tend to use the Internet everyday. Thus, if e-health is widely implanted in Brunei and reached up the whole population it would be a success as drawn from the survey that the people are able to access Internet everyday and not to mention that, they were at their 20s and working. Consequently, they afford to get an Internet access with their income received and share with their family members. Computer literacy in Brunei is enabler factor of e-health initiative, it is found that all of the respondents use computer on daily basis, while only smack percent of the respondents use computer a few times a week. This proves that most of the population is computer literate and is able to use the computer to get information. Therefore, it is believed that e-health service would be supported and acceptable by the public. Furthermore, we have already seen a very high positive response from the survey sample.

In summary, the majority of respondents are young adults, living with their family, who have completed more than high school, which explains their preference in the implementation of the systems. Not only that, they are proficient in ICT as well, considering a number of them uses the Internet and computer daily.

**Recommendations**

E-health is not going to replace any existing services in healthcare; it is extendable systems which features and services can improve quality of service. It will form healthcare services more comprehensive and reliable in providing better services. The survey confirmed respondents' expectations in healthcare are high, which create challenge for healthcare organization. In this section, we recommend new approaches and features of e-health to healthcare organization based on the finding. Figure 1 below is the recommendation which accommodates ideas of respondents about future e-health services which will be expected to increase customer satisfaction, avoid conflict, and promote better health to patient;

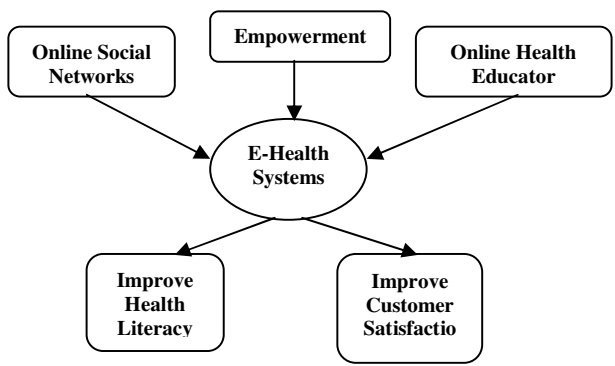

Fig. 1 e-health with features based on Survey

Empowerment is supported in the health-related literature, and it has been used in relation to customers and healthcare over the past decade. Customers or in this case patients empowerment means to support and encourage them to have more responsibility for their own health and to take a greater role in decisions about their own health care. Empowerment is a characteristic of groups and individuals that energizes them with the knowledge and confidence to act in their own behalf in a manner that best meets identified goals. In this sense, then, empowerment characterizes the manner in which patients and clinicians approach care, with mutual expectations, rights, and responsibilities. Empowerment represents a change in philosophy for both care providers and patients alike, requiring the former to abandon the authoritative control once held and the latter to assume a greater level of deliberate self involvement in the care process [19]. On an individual level, empowerment is a social process of recognizing, promoting, and enhancing peoples' abilities to meet their own needs, solve their own problems and mobilize the necessary resources in order to feel in control of their lives [11]. For consumers to be fair and equal participants in empowered partnerships with clinicians requires that they have adequate knowledge; set realistic goals; access systematic. The survey confirmed that patient preferred to have more ability and control of health information and application that concern with their health status. Making appointment online, viewing their medical records, and consulting online with medical staffs are few examples of empowerment to patient. The authorization to access health information will benefit both patient and healthcare organization. From patient perspective, their health literacy is expected to improve by the time they have better knowledge of their own health status, and for the healthcare organization is expected to be long lasting relationship since they always in need to access the service. Additionally, empowerment is also believed giving them flexible time when and where they want to look upon their health related activities. Finally, the authors suggest that the features of empowerment in e-health system should exist in four dimensions; patient-patient interaction, patient-provider interaction, disease and its treatment, and psychosocial aspect of living.

Social Networks and Social Supports are newly features introduce in the systems. Chatherine and Barbara mentioned that social networks and social supports closely affect to health outcomes [20]. Therefore, we asked respondents whether they need facilities of social networks and social supports online. Majority of respondents agreed social networks and social support online part of the e-health systems. It should be operated, managed, and maintained within healthcare's infrastructure. This is more targeted to internal patients/families within the healthcare to have conversation between patients/family within the same interest or health problem/ illness. For example, patient with diabetic would motivate to share his/her experiences, learning, and knowledge with other diabetic patients. Since patient/family who generates the contents of the Web, it can promote useful learning centre for others, not only promoting health among each others, but also it could be the best place supporting group and sharing their experiences related to all issues such as; how the healthcare does a treatment, how much it will cost them, what insurance accepted by healthcare, how is the food and nutrition provided, etc. Furthermore, it is also suggested that medical staff or doctors should be actively involved in an online forum discussion to provide professional advice. For healthcare management, conversation generates between patients in social networks site can be a starting place to construct business strategy for an organization. Social networking site should be moderated by competent healthcare staff within the managerial level to capture the conversation and listen what patients say about the service. The role of this task could campaign of healthy life for the society which is not intended for commercial benefit for short term but it is beneficial for the community.

Online Health Educator plays pivotal role in the process of interaction patient-health provider. Patients should know how to use the service. E-health services which empower patients to have better knowledge and control over their health data, yet they need to communicate with healthcare organization. In order to achieve the goal, the presence of online health educator determines success or failure of the implementation. It should be part of the implementation strategy to ensure that there are groups of staff to ensure that e-health service is managed in professional ways. Any health activities which they can accomplish through online service, they may not come physically to the hospital. Though they do not come to the healthcare centre, yet they may need an assistant or guidance from online health educator when they face difficulties interacting with systems which they need to discuss with online health educator for clarification and interpretation such as medical data, online consultation, or asking for any online service

In light of this, from the survey, confirmed that the availability of online health educator is important. Presence of online health educator is the vital point in e-health instead of ICT as tool. They are expected to have skills to interpret medical data, able to guide patient go through technical systems, and know how to respond online queries properly.

E-health promotion, education, and in emergency situations; it is believed that an e-health system would give advantages like early warning system, especially in emergency situations or pandemic. For instance, during the H1N1 flu outbreak, it is possible for health authorities to announce preventive measures over the e-health system to notify the public efficiently. In addition, this would also reduce the time-wasted in printing leaflets or posters to inform the public in such situations.

Data Ownership and External Access by Third-Parties, from our analysis, there is significant evidence suggesting that the public would prefer to keep their medical condition private and would not be willing to grant any permission to allow external parties to access such information. Therefore, our recommendation is critical to research and investigate the role of medical data ownership which data belong to patient and which parties are allowed to have limited access. We conclude that the data ownership and accessibility for patients will encourage patient to aware of their own health status which is good to improve health literacy.

The customer acceptance of technology and the demand for online healthcare services and technology among many segments of the general population has been high, there is ample evidence in the survey to support the prediction that the benefits of e-health through advanced Information and Communication Technologies have the potential to make profound impact in customer's expectation and all areas of healthcare services.

VI. CONCLUSION

Implementation of e-health will enable more patient-friendly healthcare services and at the same time improve health provider's performance. In the national level, there is also a conscious move or effort by the government to understand the needs of the public and to continuously upgrade service standards. In addition, the availability of ICT and the Internet infrastructure as well as the techno savvy of the Brunei population will assist and accelerate the implementation of e-health in the country. The general public in Brunei has responded with great support for the features and capability of e-health listed from the questionnaires. Unfortunately there is a great a gap between available health services and services expected by the public, who seem to have advance ICT literacy. From the survey also it can be concluded that most respondents agrees and are keen in having e-health services whilst requiring high degree of confidentiality on their medical information.

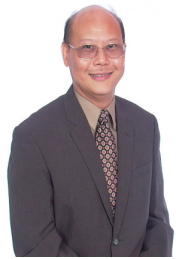

**Prof. Dr. Patrick Kim Cheng Low, Ph.D.** (South Australia), Chartered Marketer, Certified MBTI Administrator, & Certified Behavioral Consultant (IML, USA), brings with him more than 21 years of combined experience from sectors as diverse as the electronics, civil service, academia, banking, human resource development and consulting. The once Visiting Professor, Graduate School of Business, the University of Malaya (Jan to Feb 2007), Prof. Dr. Low was the Deputy Dean, Postgraduate Studies & Research, teaching in Universiti Brunei Darussalam (2009). He teaches the graduate students/ MBA in Organizational Behavior, Managing Negotiations, Leadership and Change Management, and the undergraduates in Leadership Basics, Challenging Leadership, Business and Society, Issues in Organizational Leadership, Organization Analysis & Design; and Organization Development & Change. The former Associate Dean, Director of Career Services and Chair of the Management and Marketing Department of a University in Kazakhstan (2004 to 2006) focuses on human resource management and behavioral skills training covering areas like negotiation/ influencing, leadership and behavioral modification. He can be contacted at patrick_low2003@yahoo.com.

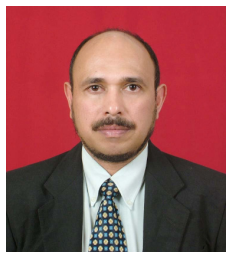

**Mohammad Nabil Almunawar** Ir. IPB Indonesia, MSc UWO Canada, Ph.D. UNSW, Australia is a senior lecturer at Faculty of Business, Economics and Policy Studies, Universiti of Brunei Darussalam (UBD), Brunei Darussalam. Dr Almunawar has published many papers in refereed journals as well as international conferences. He has many years teaching experiences and consultancies in the area computer and information systems. His overall research interest is application of IT in Management and e-commerce/e-business/e-government. He is also interested in object-oriented technology, multimedia information retrieval, health information systems and information security.

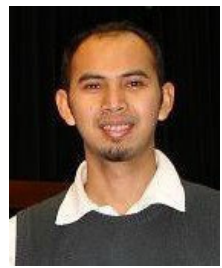

**Muhammad Anshari,** is a business information system's practitioner and researcher. He received his BMIS (Hons) from International Islamic University Malaysia, his Master of IT from James Cook University Australia, and currently, he is pursuing his PhD program at Universiti Brunei Darussalam. His professional experience started when he spent two years as an IT Business Analyst at PT. Astra International Tbk. (National Automotive Company), joined research collaboration on *Halal* Information Systems at Chulalongkorn University Bangkok - Thailand, and researcher at College of Computer and Information Science King Saud University, Riyadh-Saudi Arabia. His main research interest is on CRM, ERP, SCM, and healthcare management issues in relation with people, process and ICT. His email address is anshari@yahoo.com.

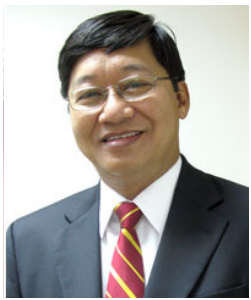

**Dr. Zaw Wint** is Senior Lecturer at PAP Rashidah Sa'adatul Bolkiah Institute of Health Sciences as well as acting dean of Graduate Studies and Research Office at Universiti Brunei Darussalam (UBD). His research domain is Developing National Science and Technology Capacity in Heart Diseases in the Country: Brunei Heart Study.